\documentclass[a4paper]{jpconf}
\usepackage{graphicx}
\begin{document}

\title{Multiple regimes and coalescence timescales for massive black hole pairs
; the critical role of galaxy formation physics}

\author{Lucio Mayer}

\address{Center for Theoretical Astrophysics and Cosmology, Institute for Computational Science, University of
Z\"urich, Winterthurestrasse 190, 8057, Z\"urich, Switzerland}

\ead{lmayer@physik.uzh.ch}

\begin{abstract}
We discuss the latest results of numerical simulations following the orbital decay
of massive black hole pairs in galaxy mergers. We highlight important differences
between gas-poor and gas-rich hosts, and between orbital evolution taking place
at high redshift as opposed to low redshift. Two effects have a huge impact and are rather 
novel in the context of massive black hole binaries. The first is the 
increase in characteristic density
of galactic nuclei of merger remnants as galaxies are 
more compact at high redshift due to the way dark halo
collapse depends on redshift. This leads naturally to hardening timescales due to 3-body encounters
that should decrease by two orders of magnitude up to $z=4$. It explains naturally the
short binary coalescence timescale, $\sim 10$ Myr, found in
novel cosmological simulations that follow binary evolution from galactic to
milliparsec scales. The second one is the inhomogeneity of the interstellar medium
in massive gas-rich disks at high redshift. In the latter star forming
clumps 1-2 orders of magnitude more massive than local Giant Molecular Clouds (GMCs)
can scatter massive black holes out of the disk plane via gravitational perturbations
and direct encounters. This renders the character of orbital decay inherently stochastic,
often increasing orbital decay timescales by as much as a Gyr. At low redshift
a similar regime is present at scales of $1-10$ pc inside Circumnuclear Gas Disks (CNDs).
In CNDs only massive black holes with masses below $10^7 M_{\odot}$
can be significantly perturbed. They decay to sub-pc separations in up to $\sim 10^8$ yr
rather than the in just a few million years  as in a smooth CND. Finally 
implications for building robust forecasts of LISA event rates are discussed.

\end{abstract}

\section{Introduction}

The orbital decay of massive black hole pairs from galactic scale separations
to distances at which gravitational wave (GW) emission takes over is governed
by a number of different physical mechanisms. In particular the decay
has a different nature in a predominantly gaseous background as opposed to a
predominantly stellar background. These two regimes are usually studied
separately, although galactic nuclei in a galaxy host might transition
from one regime to the other even multiple times in their evolution as
a result of galaxy mergers, starvation of cosmological gas accretion,
disk instabilities, and any process that can significantly affect the
gas supply and star formation rate in galactic nuclei.
The last decade has seen significant improvements in the realism of the
numerical models dedicated to tackle one or the other regime, and even
the first attempts to model both regimes in the same calculation 
(Khan et al. 2013a).

Two important general results can be drawn from the
studies of the last few years. First, it is now largely agreed that the
last parsec problem that once plagued the stellar background calculations
does not exist in a realistic galaxy host, namely one with a triaxial, rotating
stellar distribution (Berczik et al. 2006;Khan et al. 2011;2012;2013b; Vasiliev, Antonini \& Merritt 2015).
One of the most convincing demonstrations
comes from a recent analysis of orbital diffusion in the loss cone using different
types of potentials, including triaxial ones, with increasing number of particles
in  collisionless simulations (Gualandris et al. 2016). Loss cone refilling
thus appears to be effective enough in realistic galaxy hosts, allowing a successful
hardening via 3-body encounters down to separations at which gravitational
waves take over. Recently cosmological hydrodynamical 
simulations have confirmed that realistic merger remnants, in which gas dissipation
and star formation takes place, do possess the
required degree of triaxiality to keep the loss cone nearly full and drive
coalescence (Khan et al. 2016). 

The second result is that the decay in a gaseous background, once believed
to be much faster than in a stellar background
(eg Escala et al. 2005; Dotti, Colpi \& Haardt 2006; Mayer et al. 2007), turned
out to be a much more complex problem than initially suspected, with the consequence 
that there is  no simple general scenario at the moment. At variance with the
stellar background case, in which dynamical fiction and 3-body encounters
are well defined and well understood sources of drag, in
gaseous backgrounds different types of torques can dominate 
at different scales and in different stages of the orbital evolution, such as dynamical
friction, linear and nonlinear torques from a background circumnuclear rotating
disk (CND), stochastic torques from massive clouds/clumps
and large-amplitude spiral density waves, and tidal torques by 
circumbinary disks at the smallest separations (see eg Mayer 
2013, Roedig et al. 2012; Farris et al 2015).
The stochastic torque regime due to perturbations in an inhomogeneous
interstellar medium (ISM), in particular, was pointed out only recently, 
and can delay or shorten orbital decay by orders of magnitude relative
to the smooth disk case (Fiacconi et al. 2013; Mayer 2013). Finally, even in the simple case of a smooth CND
there are indications that the orbital decay may stall, or at least 
slow down significantly, once the separations of the two holes 
drops below a tenth of a parsec (Mayer et al. 2008; Chapon et al. 2013).

As a result of the multiple regimes and physical parameters of galactic nuclei 
that  can affect orbital decay, it is fair to say that black hole merger
timescales are really uncertain overall, with values ranging from
as small as 10 Myr (eg Khan et al. 2016) to as large as a few billion years (Khan et al. 2011;2012) 
depending  on the specific regime and physical conditions considered. Indeed in this
article we focus on this pivotal issue of timescales as it is highly relevant to
define the science that will be possible with LISA.

\section{Orbital decay in stellar-dominated hosts; a question of timescales}

In the last few years a number of works, both numerical and semi-numerical,
have provided evidence that there is no last-parsec problem in realistic
galactic potentials, where non-axisymmetry and some degree of rotation are
always present (Berczik et al. 2006; Vasiliev, Antonini \& Merritt 2015; 
Khan et al. 2013b; Holley-Bockelmann \& Khan 2015).
Although convergence of the hardening rates is
not yet demonstrated in collisional N-Body simulations (but see Holley-Bockelmann \& Khan
2015 on highly rotating systems), alternative approaches
such as the numerical convergence study of orbital diffusion in the loss cone 
using collisionless N-Body simulations (Gualandris et al. 2016) strongly
support the notion that the loss cone should remain nearly full in realistic
galaxy merger remnants. This in turn yields confidence in the results of 
collisional N-Body simulations of black hole binary hardening in galaxy mergers
at the currently affordable resolutions (less than $10^7$ particles, see 
eg Khan et al. 2012; Holley-Bockelmann \& Khan 2015).

\begin{figure}[h]
\includegraphics[width=34pc]{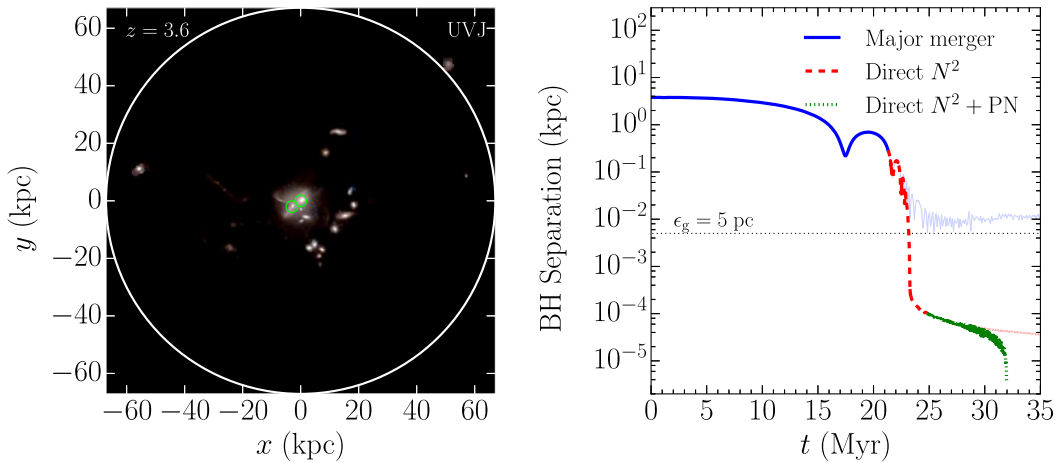}\hspace{2pc}
\caption{\label{label}On the left we show a grey-scale luminosity map (in the UVJ filter) from the ARGO cosmological
simulation showing the galaxy merger that was selected and re-simulated at much higher resolution (green circles). 
The white circle marks the virial radius of
the group-sized halo, while the green circles mark the merging galaxies. 
From the spatial scale it is clear
how the galaxies at study, which have stellar masses almost as large as those of our present-day Milky Way  ($> 10^{10}
M_{\odot}$) have sizes below 1 kpc, hence 4-5 times smaller than our Milky Way. This is expected based on the
scaling arguments described in the text. On the right we show the complete orbital separation curve for the two massive black 
holes at the center of the re-simulated merger with pc scale resolution, ending with the in-spiral driven by
GW emission (green dashed line ar 32 Myr). We plot also the results of simulations without the direct N-body
integration, that captures the hardening phase, as well as with direct N-Body but without the post-newtonian corrections
(Blanchet 2006), which
are relevant in the very last phase. We refer to Khan et al. (2016) for details.}
\end{figure}

The latter simulations find that
massive black hole binaries can merge on timescales shorter than an Hubble time.
A particularly successful case is that of ultramassive black holes in stellar dominated
hosts, for which dynamical friction is so effective to bring them all the way
to the gravitational wave emission phase without the need of a long hardening stage 
(Khan \& Holley-Bockelmann 2015). However, with this last exception, all published timescales
are really long, ranging from slightly below 1 to a few Gyr, namely they encompass a significant
fraction of the Hubble time. Note that the latter result is consistent with the simple
analytical predictions of Gould \& Rix (2000), which were assuming efficient loss 
cone refilling in a potential with the properties of present-day spheroids.

N-Body simulations mostly consider black holes at the
high mass end, $ M_{BH} > 10^7 M_{\odot}$, hence the coalescence
timescale for LISA black holes, which are much lighter, would be
even longer, assuming the same properties of the (stellar) background hold.
If merging timescales of a few Gyr are the norm
at any redshift, the consequence would be that black hole mergers would be 
exceedingly rare at $z > 2$ since the lookback time is a less than 3 Gyr then. 
This would be unfortunate since this is also the
epoch for which the discovery potential of LISA
is  very high. The inefficient coalescence in stellar backgrounds
has been and still is the motivation behind exploring 
the alternative regime in which
orbital decay is driven by gasdynamical processes. 
However, as we will illustrate in section 3, coalescence does not become necessarily
faster in gaseous backgrounds. 
Therefore it is fair to say we are  facing a {\it timescale problem} rather than
a {\it last parsec problem}.

We ought to make progress in order to obtain robust forecasts
black hole coalescence rates in the LISA band. Uncertainties in the black hole seeding
models would also generate important variations in the expected event rates (Sesana, Volonteri \& Haardt 2007). 
Yet black hole seeding models tend to yield different black hole population properties only at 
at very high redshift, while afterwards
the physics of black hole growth via gas accretion and mergers with other black holes 
become the dominant driver of the evolution of the black hole mass function ($ z > 4$, see eg Volonteri \& Natarajan 2009; 
Bonoli et al. 2014).

\subsection{Coalescence on a fast track at high redshift}

Sesana \& Khan (2015) have pointed out that the hardening timescale is determined by the
longest decay phase at sub-pc separations. This is the phase associated with the transition between dominance of 3-body
encounters and dominance of energy loss by gravitational wave emission. The two processes
have indeed rates  proportional to  $a^2$ and $a^{-3}$, respectively, where $a$ is the semi-major axis 
(hereafter separation) of the binary. By defining
a characteristic separation $a_{*/GW}$ at which the hardening rate by 3-body encounters and
the hardening rate by gravitational wave emission are equal one
obtains the lifetime of the binary at sub-pc scales as that corresponds to the
orbital decay time in the longest phase of the decay. Following Sesana \& Khan (2015).
for the latter characteristic decay timescale one finds:

\begin{equation}
t_*= { \sigma_* \over {GH \rho_* a_{*/GW}}}
\end{equation}

where $\rho_*$ and $\sigma_*$ are, respectively, the characteristic stellar density
and mean (1D) stellar velocity dispersion in the galactic nucleus,
 $H$ is an adimensional parameter expressing how close
the loss cone is to full ($F=1$ for a full loss cone, $F \sim 0.7$ for triaxial
systems employed in numerical simulations such as in Khan et al. 2012).
The dependence on density and stellar velocity dispersion is linear. However
this equation has no apparent connection with cosmology, although the parameters
that appear in it need to be informed by cosmology in modern galaxy formation
theory. Elucidating such connection is the main purpose of this section.

We start by recalling that  in the galaxy formation scenario arising in
a Universe dominated by Cold Dark Matter (CDM) cosmic structures
form bottom-up via collapse of density fluctuations combined with hierarchical merging.
The characteristic density of a dark matter halo is determined
by the virial overdensity $\rho_{vir}$.
Inside the region whose mean overdensity is the virial
overdensity a halo is virialized, namely there is a well defined equlibrium
structure for which the binding energy is set.
In the simple spherical collapse model the virial overdensity can be calculated
exactly once the cosmological parameters are known (eg Bryan \& Norman 1998).
Often for simplicity the redshift-independent
overdensity defined to be 200 times the critical density of the Universe, 
$\rho_{200}$, is adopted.
While such overdensity is only strictly valid in a Universe with matter density 
parameter $\Omega_{m} = 1$,
quantitative statements concerning redshift evolution of structures are very weakly
dependent on the presence or not of a cosmological constant (with corresponding density
parameter $\Omega_{\Lambda}$) until it begins to be dominant at $z < 1$. This is because the virial ovedensity
varies only by factors of 2 as a function of redshift relative to $\rho_{200}$ 
(Bryan \& Norman 1998).
In the remainder of this section we will  thus drop $\Omega_{\Lambda}$
since equations are simplified considerably, and afterwards we will comment on the
magnitude of the eventual corrections.

Inside a virialized halo baryons would collapse and form
a luminous galaxy. Hence, as it is customary, one
can assume that scaling relations between key galaxy structural parameters, such as
mass, radius, circular velocity (or velocity dispersion), and thus also characteristic density, 
directly reflect those of the virialized host halo because this comprises 
most of the mass, and therefore most of the binding energy
(White \& Rees 1978; Mo,Mao \& White 1998). Galaxy size will depend also on the amount of angular momentum, 
but this is also assumed to be proportional to the angular momentum of the halo and to
be approximately conserved, introducing a simple scaling factor between galaxy size
and halo virial radius. 
The virial overdensity
$\rho_{vir}$ is, by definition, a multiple of the mean density of the Universe at the
corresponding epoch, 
thereby it has to increase as $(1+z)^3$  towards high redshift. With our assumption
that $\rho_{vir}= \rho_{200}$ we are simply enforcing that the multiplication factor is
constant over time. As a consequence,
{\it if the characteristic density of galaxies follows the halo overdensity then it 
should  also scale as $(1+z)^3$.}
For instance at $z=4$ galaxies should be approximately $5^3$ times denser relative
to their $z=0$ counterparts. While this is essentially textbook structure formation
one should appreciate the very important implications in the context of massive black hole binary
coalescence. {\it Indeed based in eq. (1) this means that coalescence timescales can be 
shorter by two orders of magnitude at $z=4$ relative to $z=0$. The latter is a significant
effect.}

This simple approach neglects two aspects. The first one is that, based on the same scaling
relations between halos and galaxies, also the stellar velocity dispersion, which enters in equation (1), 
depends on redshift, as we will illustrate in a moment.
The second one is that baryonic physics can decouple to some extent
the dynamics of the baryons from that of the dark matter. Hence the scaling of halo and galaxy properties
is not necessarily homologous (although cosmological hydro simulations show it is so 
to some degree, see eg Sokolowska et al.  2016).
Indeed modern theory of  galaxy formation is essentially based on the effort to try achieve the necessary decoupling
suggested by observational constraints on galaxy properties 
via what we collectively indicate as "feedback mechanisms".  Within the same
framework, radiative cooling is also an obvious process that tends to decouple
baryons from the collisionless dark matter fluid, with a tendency to increase
density beyond the the inference of the scalings just described because important
coolants such as recombination and atomic radiative transitions increase their rate
with increasing density. These processes will shape the mass distribution in galaxies
, hence  they will determine the density profile as a function of radius.
We will return to this second point in Section 3 but for now we can continue
our reasoning as if galaxies were sufficiently  described by a characteristic density, i.e. neglecting 
their actual mass profiles.
Assuming an isothermal sphere for the velocity distribution we can
move on easily with our calculations. 
This is an oversimplification given that (1) the NFW model, that is conventionally used
to describe CDM halos, is close to the isothermal law only in the intermediate region,
and that (2) both halos and galaxies are not sppherical so they will possess significant orbital
anisotropy. Yet here we want to capture the essential notions and scalings rather than
exact numbers. Hence continuing with the isothermal model
one can write that the 1D velocity dispersion is $\sigma = V_{vir}/\sqrt{2}$, where we impose that the 
(constant) circular velocity $V_c$ is equal to the 
virial velocity $V_{vir}$ of the dark matter halo. 
The latter is automatically determined by the halo virial mass
in dissipationless spherical collapse theory once the virial overdensity and the
reference cosmological parameters are set. The virial radius is also automatically determined
in the spherical collapse model as the radius containing a region with the virial overdensity,
eg the radius at which the density equals $\rho_{200}$. The relation between halo virial mass
$M_{vir}$ and virial velocity $V_{vir}$ is (see Mo, Mao \& White 1998):

\begin{equation}
{M_{vir} = {V_{vir}^3 \over {GH(z)}}}
\end{equation}

The redshift dependence is contained in the Hubble parameter $H(z)$ which, in a generic cosmology
with non-zero cosmological constant, is defined as:

\begin{equation}
{H(z) = H_0 {[\Omega_{\Lambda, 0} + (1 - \Omega_{\Lambda, 0} - 
\Omega_{0}){(1+z)^2 + \Omega_{0}(1+z)}^3]}^{1/2}}
\end{equation}

where $\Omega_{0}$ and $\Omega_{\Lambda, 0}$ are, respectively, the total density parameter and the 
cosmological constant density parameter  at $z=0$, and $H_0$ is the Hubble constant.
Assuming  $\Omega_{\Lambda, 0} = 0$ and  $\Omega_0 = 1$ in equation (3) 
and substituting the result into equation (2) one obtains the following scaling of the 
velocity dispersion with 
redshift for a fixed halo virial mass $M_{vir}$:

\begin{equation}
{\sigma \sim V_{vir} \sim {H(z)}^{1/3} \sim {(1+z)}^{1/2}}
\end{equation}

Assuming the mass of the galaxy is simply proportional to halo mass, as strongly suggested 
by abundance matching (eg Behroozi et al. 2013), this simple scaling relation can also be 
used for  the galaxy stellar velocity dispersion as long as the galactic potential
does not modify greatly the relation between velocity and mass (which it cannot if
the galaxy and halo are in equilibrium).
At $z=4$  the relation implies an increase in velocity dispersion of  a factor of $\sim 
2.2$ relative to $z=0$ as opposed to an increase in density of more than a factor of a 100 
(see  above). 
Bringing the cosmological constant back in inside equation (3) would only change the result 
by about $6\%$. Hence, if we assume that the stellar velocity dispersion $\sigma_*$ is of 
order of the velocity dispersion of the halo, $\sigma$, we can safely  state that the main
effect as redshift increases is that the increasing density should shorten the hardening
timescale almost as fast as the density increases, namely as ${(1+z)}^3$.

The previous discussion is instrumental to understand the results of recent simulation
work. Khan et al. (2016) have made the first attempt to use a a LCDM galaxy formation simulation
as the initial condition to follow
the evolution of a massive black hole binary into the gravitational 
wave emission regime. They selected the most massive 
pair of galaxies within the high resolution sub-volume of the cosmological 
zoom-in hydrodynamical simulation ARGO (Fiacconi \& Feldmann 2015; Fiacconi, Feldmann \& Mayer 2015),
having stellar masses of a few times $10^{10} M_{\odot}$ in halos with $M_{vir} \sim 10^{12}
M_{\odot}$. Then they 
increased the resolution further in two steps using particle splitting. 
Finally they converted into stars the very little residual gas after the
major merger in order to continue the simulation as an N-body collisional
calculation with post-newtonian corrections (Blanchet 2006). Uncertainties in baryonic 
processes,especially feedback, are of course present. Nevertheless
the galaxy merger was chosen after verifying 
that it produces a galaxy whose key structural parameters match the available observational
constraints, such as the relation between stellar mass and halo mass
and the scaling relations between galaxy effective radius and its stellar velocity dispersion
(Feldmann \& Mayer 2015). In particular, the
characteristic densities and stellar velocity dispersions, which enter eq (1), match
reasonably those inferred for observed massive galaxies at high redshift
(Szomoru et al. 2012; Bezanson et al. 2013). The selected merger starts
at $z=3.6$  and is  over by $z=3.2$ (Figure 1, left panel).

\begin{figure}[h]
\includegraphics[width=18pc, angle=270]{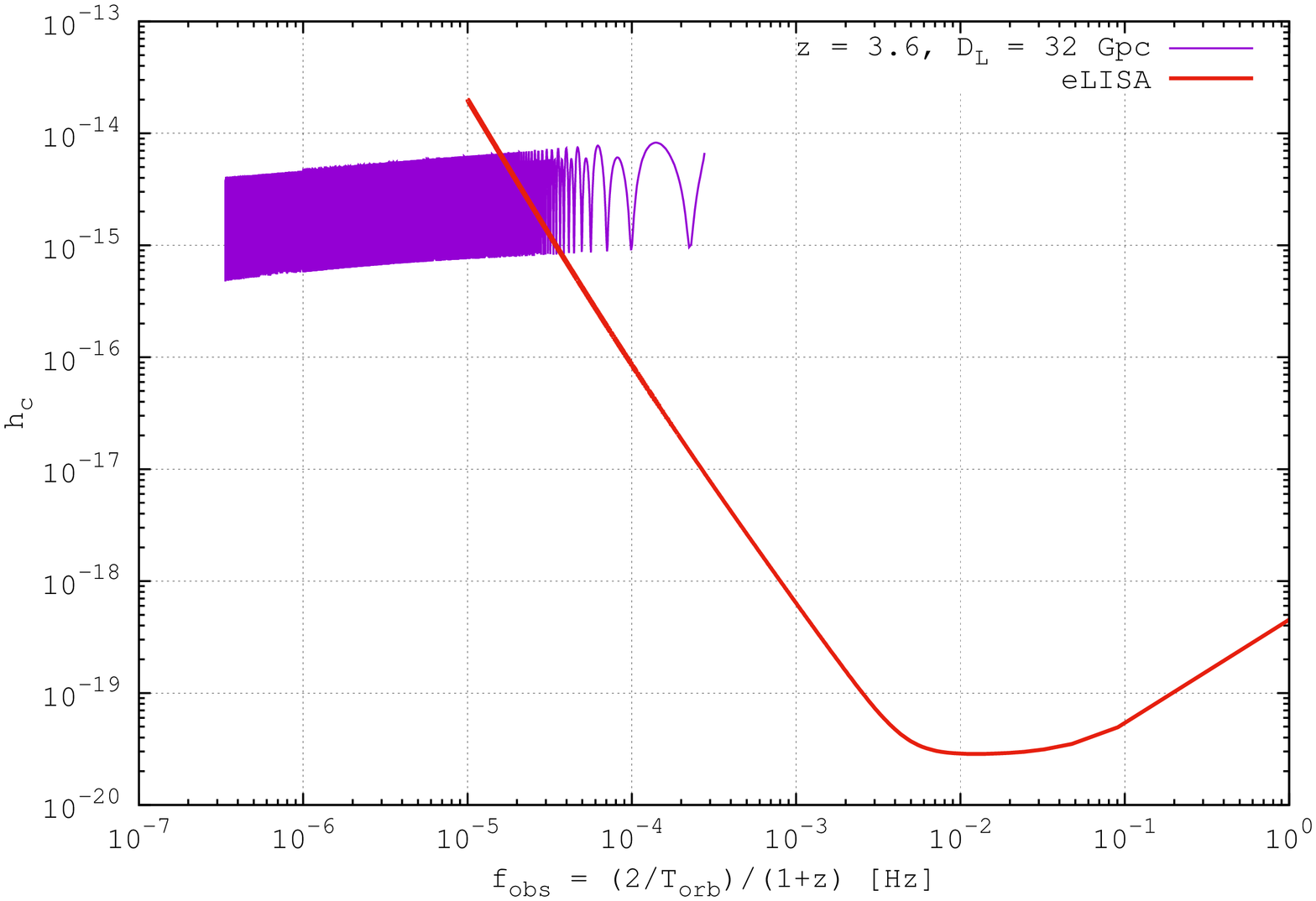}\hspace{2pc}
\caption{\label{label} Wave-form for the massive black hole merger simulated in Khan et al. (2016),
assuming a luminosity distance (32 Gpc) corresponding to $z=3.6$ in the concordance LCDM cosmology. The
LISA sensitivity curve is also shown in red, according to Klein et al. (2016).The merger should
be detected by LISA in the last phase of the in-spiral.}
\end{figure}

The timescale that we obtain for the coalescence after the galaxy merger is completed is surprisingly
small, about 10 Myr,and corresponds to the duration of the hardening phase, as the previous phase in
which the two galactic cores merge by dynamical friction inside the common halo lasts only about
a Myr (Figure 1, right panel). Can we explain such a short timescale based on the scaling 
arguments just
outlined?  The answer is yes. Indeed we attempted the rescaling procedure suggested by the above
equations in the non-cosmological models of Khan et al. (2012). Note that Khan et al (2012) used Dehnen models with
with three different inner slopes (with exponents -1.5, -1 and 0.5, respectively). We found that the
merger remnants of models with the steeper inner slope provide a better fit to the cosmological
merger remnant of Khan et al. (2016), which
reflects the effect of gas  dissipation increasing the central mass density prior to the completion
of the merger. Hence we
first rescaled those remnants to the same mass of the Khan
et al. (2016) cosmological merger remnant, and then rescaled the effective radius according to the expected 
redshift evolution so that the density was increased by a factor ${(1+z)}^3$ (the original models
were constructed to represent at $z=0$ galaxies).
Using the characteristic density of the rescaled models
in equation (1) we obtained an hardening timescale of about 30 Myr, as opposed to almost a Gyr in the
mergers using the original models adopted in Khan et al. (2012), just slightly higher than 
the timescale
we find in the cosmological simulation. This supports the notion that density scaling with redshift
is the main reason behind our short timescales as a similar result can be obtained starting with
simple idealized models once these are appropriately rescaled to account for the redshift-dependent effects
(but note that we did not scale velocity dispersion).

The very central
density and stellar velocity dispersion are the product of the balance between gas cooling and heating/stirring
by  feedback processes occurring during a nuclear starburst driven by the merger (see eg Capelo
et al. 2015 for a recent large simulation survey of gas-rich galaxy mergers). 
The absence of black hole accretion and AGN feedback in the simulations of Khan et al. (2016) should not be 
a worry because we verified the local relation between black hole mass and central stellar velocity 
dispersion is reproduced. Furthermore, as already mentioned, central velocity dispersions are reasonable 
when  compared to observed  massive galaxies at $z > 2$ with similar luminosities. 
As there is mounting evidence that, at high z, black hole masses are shifted upwards relative 
to the relation, probably suggesting a lesser role of AGN feedback in regulating the
concurrent growth of host galaxy and black hole masses
(Trakhtebroth et al. 2015), the black hole masses chosen in these simulation are probably on 
the conservative side.

In lower mass galaxies, in which feedback has a stronger effect due to their shallower potential
well, it is conceivable  that the characteristic
nuclear density might be reduced as a result of the starburst, an effect that has indeed been studied
extensively for dwarf galaxies in cosmological hydrodynamical simulations. In this scale
the evolution of the characteristic baryonic density with redshift might depart from the
simple scaling arguments based on the dark halo evolution.
In progenitors of Milky 
Way-sized galaxies, whose mass would be an order of magnitude lower than the current mass of the
Milky Way at $z=0$ at $z \sim 2-3$ (see eg Guedes et al. 2011;Agertz \& Kravtsov 2015), the conditions 
in the host might be just intermediate.
The study of black hole growth of a Milky Way-like galaxy carried out by
Bonoli et al. (2016) using the ErisBH simulation does indeed reveal that nuclear gas inflows are rather weak even
at $z \sim 2-4$, when the last significant mergers occur. This reflects the fact that feedback processes
are very effective at counteracting radiative cooling in this case. Since such 
Milky-Way progenitors would likely host black holes in the range $10^5-10^6 M_{\odot}$, the typical
target of LISA (Amaro-Seoane et al. 2013), massive black hole binary evolution in this type of galaxies should be 
studied  thoroughly in the future.

\section{The perilious journey of BH pairs in clumpy gaseous 
disks; from galactic to circumnuclear regions}

The previous section focused on the sinking and coalescence of supermassive black holes 
($M > 10^7 M_{\odot}$) in massive galaxies, a category
of objects that will be only marginally in the detection window of LISA. In the early phase
of their in-spiral they would, however, provide a  very
strong signal, with strong signal-to-noise ratio, at relatively low redshifts ($z < 4$) as 
in the case of the system studied by Khan et al. (2016) (see Fig.2).

Furthermore, the results presented in the last section are mostly relevant to gas-poor galaxies.
The galaxy host merger remnant in Khan et al. (2016) indeed belongs
to the category of galaxies that leave the star forming main sequence and become "red and 
dead" relatively early, at $z > 2$, such as the red nuggets identified in the CANDELS survey
(Barro et al. 2013). While it is generally believed that more massive galaxies quench their 
star formation earlier,  galaxy evolution is a nonlinear process which
occurs along multiple tracks, hence even among massive galaxies star formation may continue
or even peak at an epoch later than that of the merger remnant of Khan et al. (2016).
Indeed at $z \sim 2-3$ we witness the existence of massive star forming galaxies that
are very clumpy and disky (Tacconi et al. 2013; Forster-Schreiber et al. 2011, Wisnioski et al. 2015).
This  has often interpreted as evidence for massive, self-gravitating 
gas disks that fragment due to a Toomre-like instability (Bournaud et al. 
2010; Ceverino et al. 2010). While the 
quantitative aspects of gravitational instability of massive gas disks at high-z are 
currently debated and may strongly depend on feedback processes 
(Tamburello et al. 2015; Mayer et al. 2016), it is important to ask how would 
massive black hole pairs evolve and sink in such clumpy disks .
In particular, what would be the typical timescales for massive black hole coalescence 
compared to those in relatively smooth galactic disks at low redshift, or to those in 
gas-poor galaxies dominated by the stellar background physics?

\begin{figure}[h]
\includegraphics[width=19pc]{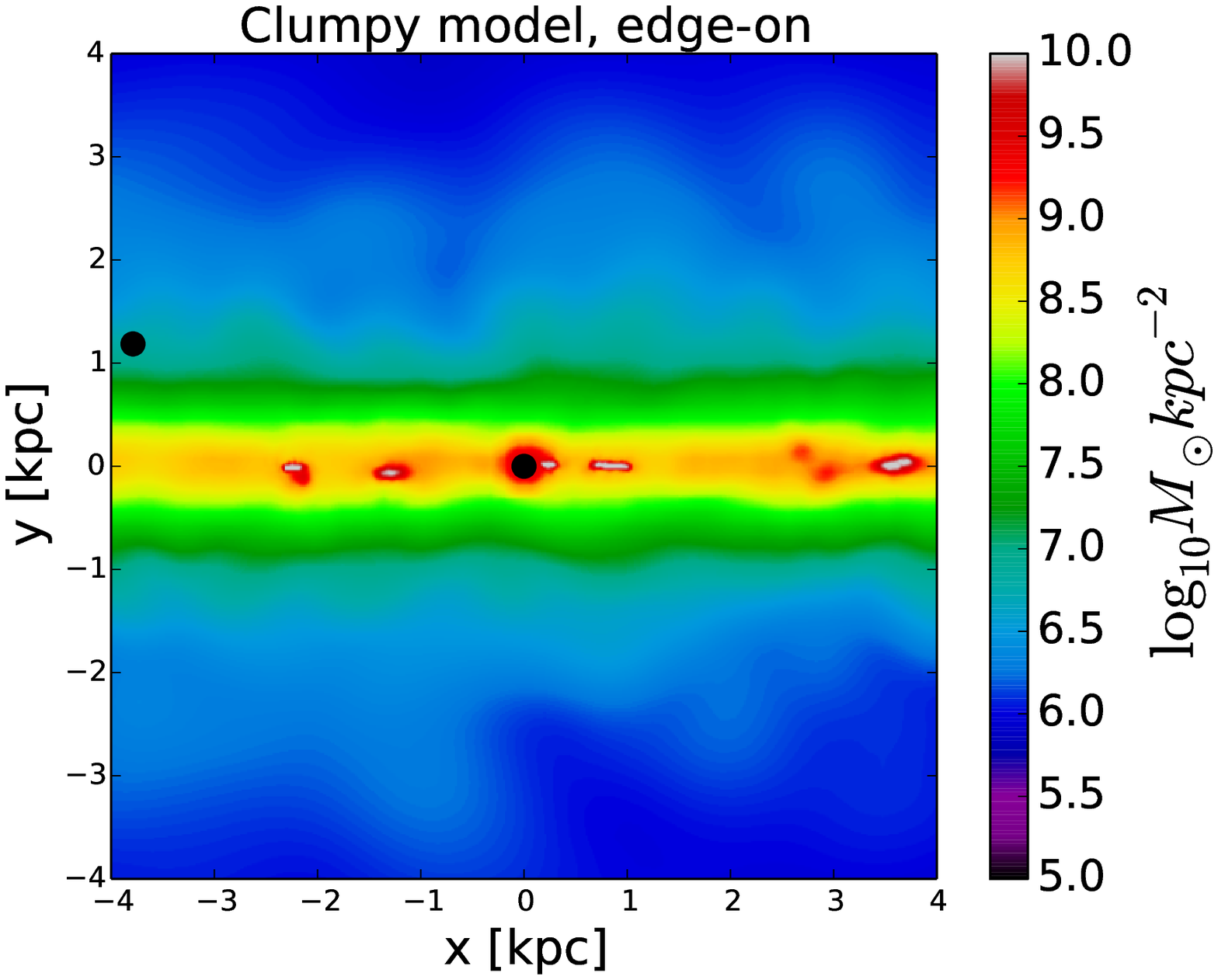}\hspace{2pc}
\includegraphics[width=19pc]{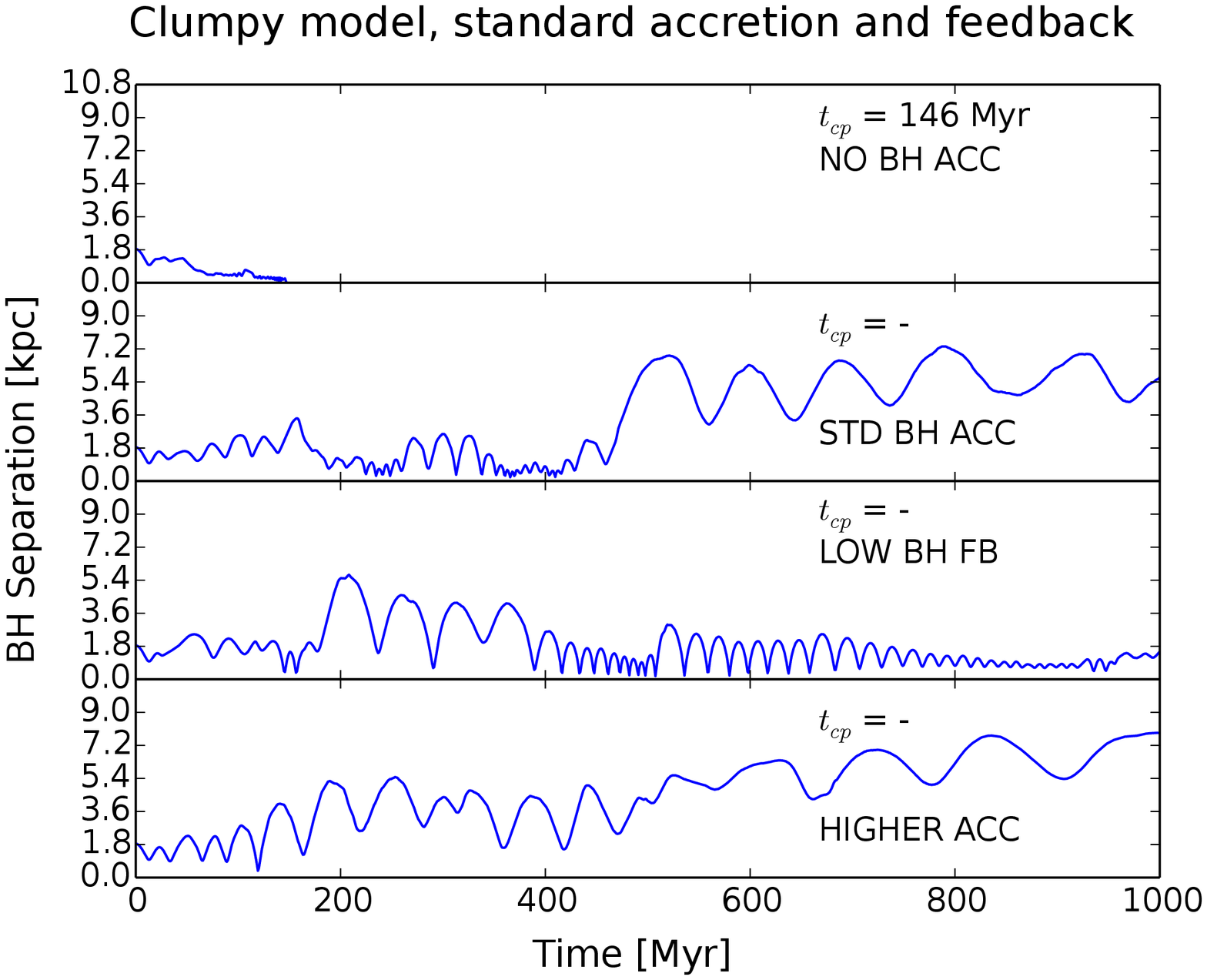}\hspace{2pc}
\caption{\label{label}On the left we an edge-on color-coded density map of one of the clumpy hi-z galaxies from
Tamburello et al. (2017), showing the ejection of the secondary black hole out of the disk plane (black dot on top
left corner of panel). On the right
we show a few examples of how the orbital separation of the secondary BH relative to the primary evolves with time.
All runs shown have BH accretion and feedback except the one used in the top panel, which only includes BH accretion.
Parameters related to the details of the accretion and feedback sub-grid models differ in the various runs (see
labels in Figure and Tamburello et al. 2017).
As it is seen there is a high degree of stochasticity in the orbital separation evolution, but in most cases the
secondary keeps wandering at large distances even after a billion years from the beginning of the simulation (but
see Tamburello et al. (2017) on the effect of a massive extended stellar bulge).}
\end{figure}

In Tamburello et al. (2017) we considered the decay of two massive black holes that are 
already embedded
in a gas-rich massive galactic disks subject to fragmentation into massive star forming clumps. 
We considered also the effect of black hole
accretion and feedback, as well as star formation and feedback from supernovae explosions. We
explored a large suite of simulations with galactic hosts having different masses and gas fractions,
and choosing different eccentricities for the orbit of the secondary black hole (the mass ratio is fixed
of the two black holes to 10:1, which should be statistically representative as the same mass ratio 
is most typical for halo/galaxy mergers).  The main result is that the lighter black hole, 
with mass in the range $4 \times 10^7 - 10^8 M_{\odot}$ ($1/5$ of the mass of the primary)
experiences repeated gravitational scattering by the
most massive clumps as well as by strong spiral density waves often 
This often results in its ejection from
the disk plane, slowing down its orbital decay by at least an order of magnitude as dynamical 
friction is  suppressed in the low density
envelope around disk midplane (Fig. 3).
In a companion paper
(Tamburello et al. 2016) we have shown how such clumps do reproduce the properties of observed star
forming clumps in high-z massive galaxies once appropriate corrections for observational biases
(limited sensitivity and resolution) are taken into account.

The suppression of orbital decay is exacerbated when AGN feedback is included as both dynamical friction
and disk torques are weaker even when the secondary is still in the disk plane due to local heating of the
gaseous background. Since it sinks slower, the secondary can be ejected when it is further away from 
the primary in this case.
The secondary, following an ejection, can still wander at distances greater than 1 kpc more after 1 Gyr.
decay has started. Taken at face value this would imply a possible 
complete abortion of the coalescence process. However, since Tamburello et al used galaxy models that lack the 
an extended bulge/spheroid component, dynamical friction was underestimated away from the
galactic disk plane.
By computing analytically the extra drag resulting from an Hernquist
spheroid with realistic structural parameters, it was found that, if no more ejections occur,
, wandering secondaries should coalesce on timescales of order 1 billion year. 
This is still a very long timescale,
comparable, if not longer, than the typical merging timescales of galaxies at $z > 2$.

The clumpiness of the ISM is an important factor even for the late stage of black hole pair decay,
at 10 pc scales and below, near the phase in which a bound binary forms. 
This applies to gas-rich galactic hosts at any redshift as circumuclear disks (CNDs) of gas and stars are ubiquitous
at low z in merger remnants as well as in the central regions of Seyfert galaxies (Medling et al. 2014;
Izumi et al. 2016) , which structurally 
typical late-type spiral galaxies. Fiacconi et al
(2013) carried out the first study of black hole pairs in CNDs, documenting the
dramatic effect of gravitational scattering by clumps and spiral density waves that we just 
illustrated. They considered black holes right in the LISA window ($10^5- 10^7 M_{\odot}$).
They also found that in some cases the decay can be accelarated due to the capture of
a massive gas cloud ($M > 10^6 M_{\odot}$) by the secondary.
They proposed the general notion that orbital  decay is {\it 
stochastic}  in a clumpy CND, with orbital decay timescales to $0.5$ pc separations ranging from a few Myr to 
as much as 100 Myr.
The shortest timescales agree with previous results for smooth
CNDs (see Escala et al. 2005 , Dotti et al. 2006;2007, Mayer 2013). Statistically, though, a largest fraction
of the simulations resulted in long timescales, in line with the same findings of Tamburello et al. (2017),
especially for black holes on eccentric orbits. 

These simulations were quite idealized in the physical treatment, lacking feedback mechanisms in particular,
which should play a major role in CND, regions that are often subject to starbursts.
Del Valle et al. (2015) and Souza Lima et al. (2016) recently carried out a similar studies but included a much richer 
inventory of physical processes. Del Valle et al. (2015) included star formation and a weak form of SN feedback,
while Souza Lima et al. (2016) employed a popular effective sub-grid model of SN feedback, the blastwave
feedback (Stinson et al. 2006), which is known to produce realistic stellar masses and disk sizes in galaxy
formation simulations (Guedes et al 2011), and even added black hole accretion and AGN feedback  
in the same way as Tamburello et al. (2017).
They confirmed the stochastic orbital decay picture and the prevalence of suppressed orbital decay with
timescales in the range 50-100 Myr. 
The same long orbital decay timescales were found in a complimentary study modeling the
entire massive black pair evolution before and after a major galaxy merger. This study also addresses
the formation of the circumnuclear disk after the merger, but was restricted to only
one initial condition for the galaxy merger due to the high computational burden introduced by deep the multi-scale 
nature  of the calculation (Roskar et al. 2015). 
Souza Lima et al. (2016) is the only work, however, that includes also the effect of BH accretion and AGN
feedback at CND-scales.
Thanks to that they uncovered a new process, dubbed "wake evacuation effect", by which the
secondary black hole carves a hole in the outer, lower density region of the CND, as it launches a hot
pressurized bubble resulting from AGN feedback (Figure 5).
The result is that dynamical friction
and disk torques are stifled as the  CND-black hole dynamical coupling is temporarily stifled by the
presence of a large cavity. As a result, the secondary decays slower even before a strong interaction
with a clump or spiral wave occurs, often resulting in an ejection when it is still far from the
center. 
In the runs with AGN feedback the outcome is thus an even longer delay, of order a Gyr , which however
should be regarded as an upper limit in absence of an extended massive stellar spheroid, for the same
argument made in Tamburello et al. 2017 in the case of high-z galactic disks. When dynamical
friction by an extended  spheroid is added, for realistic parameters we obtained a decay timescale
(to fraction of parsec separation, namely close to the resolution limit) of order a 
a few $10^8$ yr. This
is still almost two orders of magnitude longer than the decay timescale in a smooth CND and
no feedback processes included (see eg Mayer 2013).
These effects are only some of those that are possible in a complex multi-phase CND. For example, recently
Park \& Bogdanovic (2017) have shown that radiation pressure can also produce an extra drag 
pulling opposite to the dynamical friction wake, which again slows down the orbital decay.

In summary, the presence of a clumpy ISM as well as the concurrent effect of feedback processes complicate
significantly the picture in gas-rich systems, suggesting that timescales in the range $10^8-10^9$ 
yr are
likely for the decay to sub-pc separations from galactic scales, with the longer timescales more likely 
in the high-z 
clumpy galaxy phase. Despite uncertainties it is clear that, at high z, such timescales are much longer than those
in gas-poor galaxies in which the decay is governed by the physics of the stellar dynamical background. At low z, 
instead, they turn out to be comparable or shorter than those in purely stellar backgrounds , as the simulations
of Khan et al. (2012) found a typical timescale of order a Gyr for the hardening phase by 3-body encounters.

\section{Summary and Outlook}

The latest developments in modeling the orbital decay of massive black hole binaries that we just illustrated
have turned upside-down the conventional notion that orbital decay of massive black hole binaries should be faster and
more efficient in gas-rich galaxies relative to gas-poor galaxies. In 
particular at high reshift, while simulations finally show fast 
coalescence for massive black 
holes in gas-poor merger remnants of massive galaxies, with timescales at least an order of magnitude 
shorter than the merging time of galaxies themselves ($10^7$ yr as opposed to $10^8-10^9$ yr), the opposite
seems to be the case for similarly massive gas-rich disks due to the perturbing action of the
their inhomogeneous, clumpy interstellar medium. The existence of such clumpy medium is an inevitable
consequence of the self-gravity of massive gas disks, although detailed 
properties such as the mass spectrum of cold
clouds depend on the complicated interplay between gas physics and feedback processes.
In these systems orbital decay timescales acquire a stochastic character, with
values ranging from $10^8$ yr to $> 1$ Gyr, namely longer than the galaxy 
merging timescales at $z > 1$.
These estimates still do no account for the final
phase of decay below $0.1$ pc due to the resolution limitations of current multi-scale hydrodynamical simulations.

A word of caution must be spent in that there is still significant uncertainty in how to model feedback processes
in galaxy formation and evolution, which in turn has a major impact on the nature of the ISM. The latest
generation of strong feedback models, which are favoured by some constraints available on the rate of
assembly of stellar masses in galaxies, do reduce significantly the clumpiness of the ISM even in the
most massive high redshift disks (Mayer et al. 2016; Oklopcic et al. 2016). While massive star forming clumps
are observed in massive disks at high-z, their actual nature is debated, especially whether they are bound
coherent structures or the collection of much smaller, weakly bound structures (Behrendt et 
al. 2016;
Tamburello et al. 2017). This of course will have an impact on how strong are the dynamical perturbations
that disk substructure can induce on the orbits of massive black holes.

At low z it is established and inevitable that the ISM is clumpy at the scale of Giant Molecular Clouds
(GMCs), namely at the scale of $10^6-10^7 M_{\odot}$ (the largest cloud masses are observed in galactic nuclei,
eg in our Galactic Center, see Oka et al. 2001).
These cloud masses appear to be large enough to cause significant perturbations to decaying massive black holes 
with masses best accessible to LISA, namely $< 10^7 M_{\odot}$, and their scattering action
delays decay, in most cases leading to timescales $> 
10^8$ yr to reach sub-pc scale separations (Souza Lima
et al. 2016). More massive black holes would be unaffected as shown by simple analytical arguments
(Fiacconi et al. 2013). New effects caused by AGN feedback, such as wake evacuation or anisotropic
radiation pressure, can stifle the orbital
decay further (Souza Lima et al. 2016; Park \& Bogdanovic 2017).

The picture that is emerging is that the orbital decay of massive black holes is tightly linked to the
properties of their galactic hosts, and to how they evolve in redshift. It has never been so evident
as it is now that understanding the evolution of massive black holes requires control of the 
background
galaxy formation theory, which currently is developed primarily with supercomputer simulations of
galaxy formation in a cosmological context. This has two important implications, both crucial
in preparation for LISA
The first is that the potential of gravitational wave detection from massive black holes as a new
probe of cosmic structure formation is much higher than we could ever suspect. The second is that
to fully exploit such potential we need to devote significant resources to supercomputer simulations
addressing all possible regimes of orbital decay, from large to small scales, and ranging across
a wide variety of possible galaxy hosts, especially at $z > 1$ and for massive black holes
below $10^8 M_{\odot}$, the regime where the discovery potential of LISA is highest.

In essence the latest simulations reveal a high complexity of massive black hole orbital decay, with
multiple regimes and multiple physical effects involved. It is now clear that semi-analytical models
describing evolving black hole populations in hierarchical galaxy formation, the tool that has been used
so far to make predictions for LISA detections (eg Barausse 2012) need the input of such simulations
to make reliable forecasts. Without that black hole merger rates could be off by orders of magnitude. The
same applies to recent models coupling recipes for black hole orbital decay with large scale cosmological
hydro simulation that cannot resolve better than kpc scales (Salcido et al. 2016).A joint effort
of the different communities connected to the LISA science goals seems necessary to fill this gap and
make future progress.

\subsection{Acknowledgments}
The author thanks his many collaborators, in particular Fazeel Khan, Pedro Capelo, Rok Roskar, Andreas Just, 
Peter Berczik, Silvia Bonoli, Romain Teyssier \& Jillian Bellovary. A special thank goes to current and former PhD 
students, who have carried out a large part of the research described in this article, in particular Simone Callegari, Davide Fiacconi,
Valentina Tamburello \& Rafael Souza Lima. He acknowledges stimulating discussions over last 
years with Monica Colpi, Piero Madau, 
Tamara Bogdanovic, Alberto Sesana, Scott Tremaine, Zoltan Haiman, Massimo Dotti, Andres 
Escala \& Alessia Gualandris.

\section*{References}



Amaro-Seoane, P., Aoudia, S., Babak, S., et al. 2013, {\it GW Notes}, {\bf 6}, 4 

Agertz, O. \& Kravtsov, A. V., 2015, {\it ApJ}, {\bf 804}, 18

Begelman, M. C.,  Blandford, R. D., \& Rees, M. J. 1980,  {\it Nature}, {\bf 287}, 307

Behrendt M., Burkert A. \& Schartmann M., 2016, ApJ, {\bf 819}, L2

Barro, G., et al. 2013, {\it {\it ApJ}}  {\bf 765}, 104

Barausse, E., 2012, {\it MNRAS}, {\bf 423}, 2533

Berczik, P., Merritt, D., Spurzem, R., \& Bischof, H.-P. 2006, {\it ApJL}, {\bf 642}, L21

Behroozi P. S., Wechsler R. H., Conroy C., 2013, {\it ApJ}, {\bf 770}, 57

Bezanson, R., van Dokkum, P. G., Tal, T., et al. 2009, {\it ApJ}, 697, 1290 Bezanson, R., van Dokkum, P. G., van de Sande, 
J.,  et al. 2013, {\it ApJ}L, {\bf 779}, L21

 Blanchet, L. 2006, {\it LRR}, {\bf 9}, 4

  Bonoli, S., Mayer, L., \& Callegari, S. 2014, {\it MNRAS}, {\bf 437}, 1576

  Bonoli, S., Mayer, L., Kazantzidis, S., Madau, P., Bellovary, J., Governato, F., 2016, {\it MNRAS}, {\bf 459}, 2603

  Bournaud F., Elmegreen B. G., Teyssier R., Block D. L., Puerari I., 2010,
{\it MNRAS}, {\bf 409}, 1088

  Bryan, G.L. \& Norman, M.L., 1998, {\it ApJ}, {\bf 495}, 80

  Capelo, P. R., Volonteri, M., Dotti, M., et al. 2015, {\it MNRAS}, {\bf 447}, 2123

  Ceverino D., Dekel A., Bournaud F., 2010, {\it MNRAS}, {\bf 404}, 2151

 Chapon, D., Mayer, L., \& Teyssier, R. 2013, {\it MNRAS}, {\bf 429}, 3114

  del Valle, L., Escala, A., Maureira-Fredes, C., et al. 2015, {\it ApJ}, {\bf 811}, 59

  Dotti, M., Colpi, M. \&  Haardt, F., 2006, {\it MNRAS}, {\bf 367}, 103

 Dotti, M., Colpi, M., Haardt, F., \& Mayer, L. 2007, {\it MNRAS}, {\bf 379}, 956

  Escala, A., Larson, R. B., Coppi, P. S. \& Mardones, D., 2005, {\it ApJ}, {\bf 630}, 152

  Farris, B., Duffel, P., MacFadyen, A,m \& Haiman, Z., 2015, {\it MNRAS}, {\bf 447}, L80

  Feldmann, R. \& Mayer, L., 2015, {\it MNRAS}, {\bf 446}, 1939

 Ferrarese, L., \& Merritt, D. 2000, {\it ApJ}L, {\bf 539}, L9

 Fiacconi, D., Feldmann, R., \& Mayer, L. 2015, {\it MNRAS}, {\bf 446}, 1957
  
  Fiacconi, D., Mayer, L., Roskar, R., \& Colpi, M.  2013,
{\it ApJ}L, {\bf 777}, L14

  Gould, A., \& Rix, H.W, 2000, {\it ApJ}L, {\bf 532}, L29

Graham, M. J., Djorgovski, S. G., Stern, D., et al. 2015, {\it Nature},
{\bf 518}, 74 Guedes, J., Callegari, S., Madau, P., \&
Mayer,
L. 2011, {\it ApJ}, {\bf 742}, 76

Gualandris, A., Read, J. I., Dehnen, W., \& Bortolas, E. 2016, {\it MNRAS}, {\bf 463}, 2301

  Guedes, J., Callegari, S., Madau, P. \& Mayer, L., 2011, {\it ApJ}, {\bf 742}, 76

  Holley-Bockelmann K., Khan F. M., 2015, {\it ApJ}, {\bf 810}, 139

  Izumi, T., Kawakatu, N., \& Kohno, K. 2016, {\it ApJ}, {\bf 827}, 81

  Khan F. M., Just A., Merritt D., 2011, {\it ApJ}, {\bf 732}, 89

  Khan F. M., Preto M., Berczik P., Berentzen I., Just A., Spurzem R., 2012,
{\it ApJ}, {\bf 749}, 147

  Khan F. M., Holley-Bockelmann K., Berczik P., Just A., 2013a, {\it ApJ}, {\bf 773},
100

   Khan, F.M., Fiacconi, D., Mayer, L., Berczik, P., \& Just, A., 2013b, {\it ApJ}, {\bf 828}, 73

  Khan, F.M.,  Holley-Bockelmann, K., \& Berczik, P., 2015, {\it ApJ}, {\bf 798}, 103

  Klein, A., et al. 2016, {\it PhysRevD}, {\bf 93}, 2, id.0240004

  Mayer, L. 2013, {\it Classical and Quantum Gravity}, {\bf 30}, 244008

  Mayer, L., Kazantzidis, S., Madau, P., et al. 2007, {\it Science}, {\bf 
316}, 1874

 Mayer L., Tamburello V., Lupi A., Keller B., Wadsley J., Madau P., 2016,
{\it ApJ}, {\bf 830}, L13

  Medling, A. M., U., Vivian;, Guedes, J., Max, C. E., Mayer, L., Armus, L., Holden, B.,  Rokar, R., \& Sanders, David, 2014,
{\it ApJ}, {\bf 784}, 70

  Merloni, A., Bongiorno, A., Bolzonella, M., et al. 2010, {\it ApJ}, {\bf 708}, 
137

  Milosavljevic, M., \& Merritt, D. 2001, {\it ApJ}, {\bf 563}, 34

  Mo, H.J, Mao, S. \& White, S.D.M., 1998, {\it MNRAS}, {\bf 295}, 319

  Oka, T., Hasegawa, T.,  Sato, F.,  Tsuboi, M., Miyazaki, A, \&  
Sugimoto, M., 2001, {\it ApJ}, {\bf 562}, 348

  Oklopcic A., Hopkins P. F., Feldmann R., Keres D., Faucher-Giguere C.-A.,
Murray N., 2016, preprint, (arXiv:1603.03778)

  Park, K. \& Bogdanovic, T. 2017, submitted to {\it ApJ},  
(2017arXiv170100526P)

  Roedig, C., Sesana, A., Dotti, M., Cuadra, J., Amaro-Seoane, P., Haardt, 
F. 2012, A\&A, {\bf 545}, 127

 Roskar, R., Fiacconi, D., Mayer, L., et al. 2015, {\it MNRAS}, {\bf 449}, 494

 Sesana, A., \& Khan, F. M. 2015, {\it MNRAS}, {\bf 454}, L66

  Stewart, K. R., Bullock, J. S., Barton, E. J., \& Wechsler, R. 
H. 2009, {\it ApJ},
{\bf 702}, 1005

  Sesana, A., Volonteri, M., \& Haardt, F., 2007, {\it MNRAS}, {\bf 377}, 1711

  Salcido, J., Bower, R., et al. 2016, {\it MNRAS}, {\bf 463}, 870

  Stinson, G., Seth, A., Katz, N., et al. 2006, {\it MNRAS}, {\bf 373}, 1074

  Souza Lima, R., Mayer, L., Capelo, \& Bellovary, J., 2016, {\it ApJ}, in press (arXiv:1610.01600)

Szomoru, D., Franx, M., \& van Dokkum, P. G. 2012, {\it ApJ}, {\bf 749}, 121

   Tacconi L. J., et al., 2013, {\it ApJ}, {\bf 768}, 74

  Tamburello V., Mayer L., Shen S., Wadsley J., 2015, {\it MNRAS}, {\bf 453}, 
2490

  Tamburello, V. Rahmati, A., Mayer, L., Cava, A., Dessauges-Zavadsky, M., \& Schaerer, D., 2016,  {\it MNRAS}, 
submitted, (arXiv:1610.05304)

  Tamburello, V., Capelo, P. R., Mayer, L.,  Bellovary, J. M.,  Wadsley, 
J. W. 2017, {\it MNRAS}, {\bf 464}, 2952

Trakhtenbrot, B., Urry, C. M., Civano, F., et al. 2015, {\it Sci}, 
{\bf 349}, 168

  Vasiliev, E., Antonini, F., \& Merritt, D. 2015, {\it ApJ}, 810, 49

  Volonteri, M. \& Natarajan, P., 2009, {\it MNRAS}, {\bf 400}, 1911

  White, S.D.M. \& Rees, M.J., 1978, {\it {\it MNRAS}   }, {\bf 183}, 341

Wadsley, J. W., Stadel, J., \& Quinn, T. 2004, {\it NewA}, {\bf 9}, 137

Wisnioski, E., Forster Schreiber, N. M., Wuyts, S., et al. 2015, {\it 
{\it ApJ}}, {\bf 799}, 209


\end{document}